\documentclass{article}

\def\p{\partial}
\def\e{{\rm e}}
\def\d{{\rm d}}
\def\ie{i.e. }
\def\eg{e.g. }
\def\etal{et al. }

\def\rsh{r_{\rm sh}}
\def\stara{{\rm cool}}
\def\rns{r_{\stara}}

\def\M{{\cal M}}

\begin{document}
\title{Non-radial instability of stalled accretion shocks: advective-acoustic cycle}

\author{Thierry Foglizzo \& Pascal Galletti \\
{\it Service d'Astrophysique, CEA-Saclay,} \\
{\it 91191 Gif-sur-Yvette, France} }
\date{Austin, Texas, June 10-13, 2003}

\maketitle

\abstract{The linear stability of stalled accretion shocks is investigated
in the context of core collapse of type II supernovae.
We focus on a particular instability mechanism based on the
coupling of acoustic perturbations with advected ones (vorticity,
entropy). This advective-acoustic cycle takes place between the
shock and the nascent neutron star.
Both adiabatic and non-adiabatic processes may contribute to this
coupling, but only adiabatic ones are considered in this first
approach.
The growth time of the adiabatic instability scales like the
advection time, and is dominated by low degree modes l=0,1,2.
Non radial modes (l=1,2) found unstable by Blondin et al. (2003)
can be related to this mechanism.}

\section{Introduction}
Shocked accretion onto the surface of a compact star is known to be unstable in the context of magnetized white dwarfs, leading to shock oscillations (from Langer, Chanmugam \& Shaviv 1981, hereafter LCS81, to Saxton \& Wu 2001). 
Houck \& Chevalier (1992, hereafter HC92) made a linear stability analysis of shocked accretion onto a neutron star, and found an instability reminiscent of the instability found by 
LCS81. HC92 showed specific cases where the cooling occurs mostly in a thin layer at the surface of the neutron star, while the flow is essentially adiabatic above it. The mechanism of the instability was described by LCS81 and subsequent authors as a kind of thermal instability: if the shock surface is moving outwards, the higher incident velocity in the frame of the shock produces a higher temperature blob, which pushes the shock further out if the increased cooling time exceeds the increased advection time. This cycle, however, resembles the unstable adiabatic cycle described by Foglizzo \& Tagger (2000, hereafter FT00), and Foglizzo (2001, 2002, hereafter F01,F02) in the context of shocked spherical accretion onto a black hole. In this case the acoustic feedback is purely adiabatic and is due to the advection of vortical/entropic perturbations from the shock to the accretor. A similar coupling must also take place if the accretor is a neutron star. Is the instability found by HC92 due to a cooling process, in the spirit of LCS81, or is there a significant contribution of the adiabatic coupling between vortical/entropic and acoustic perturbations ?
The recent work of Blondin, Mezzacappa \& DeMarino (2003, hereafter BMD03) seems to support this second hypothesis. However, is the acoustic feedback identified by BMD03 linear or due to turbulence ?
Understanding the physical mechanism underlying this instability could prove useful in order to evaluate its role when realistic non-adiabatic processes are taken into account.

\section{The perturbed adiabatic flow viewed as a forced oscillator\label{sectmethod}}

In order to distinguish between adiabatic processes and non adiabatic ones, the flow structure between the shock $\rsh$ and the surface of the nascent neutron star $r_\star$ is schematized as an adiabatic flow above a cooling layer $\rns$. The present study focuses on the stability of the adiabatic part of the flow $\rns<r<\rsh$. 
The entropy and vorticity equations can be integrated 
explicitly as in F01. The Rankine-Hugoniot conditions at the shock impose that entropy and vorticity  perturbations $\delta S,\delta w$ are simply related through:
\begin{eqnarray}
(\delta w_{r},\delta w_{\theta},\delta w_{\varphi})=\left(0,-{c^{2}\over \gamma rv\sin\theta}{\p\delta S\over\p\varphi},{c^{2}\over \gamma rv}{\p\delta S\over\p\theta}\right).
\end{eqnarray}
The differential system satisfied by perturbations is the same as in F01 (Eqs.~(B18-B19)), only the functions $\M, v,c$ describing the stationary flow are different. 
Pressure perturbations satisfy a differential equation with a source term due to entropy/vorticity perturbations (Eq.~(4) of F01) :
\begin{eqnarray}
\left\lbrace{\p^2\over\p r^2}+a_1{\p\over\p r}+a_0\right\rbrace{\delta p\over p}
={-\Delta\over v^2(1-\M^2)}
{\delta p_S\over p}.\label{diffpress}
\end{eqnarray}
The source term $\delta p_S$ is:
\begin{eqnarray}
{\delta p_S\over p}&\equiv&-{1\over c^2} {v\over i\omega}{\p\over\p r}
\left\lbrace
\left(1-{2v\over i\omega}{\p\log\M\over\p r}\right){\omega^2c^2\over\Delta}\right\rbrace
\delta S,\label{pS}\\
\Delta&\equiv&\omega^2+l(l+1){v^2\over r^2}+2i\omega{\p v\over \p r},\label{defdelta}.
\end{eqnarray}
Eq.~(\ref{pS}) thus describes the local production of pressure perturbations due to advection of entropy and vorticity perturbations in a inhomogeneous flow. It characterizes the "excitator", whereas the left hand side of Eq.~(\ref{diffpress})  characterizes the "oscillator".
According to Eq.~(\ref{pS}), the local strength of the excitator depends on the value of the ratio 
$\omega r/v$. 
\begin{eqnarray}
{\Delta\over v^2}\left({\delta p_S\over p}\right)_{{\omega r\over v}\gg 1}&\sim&-{1\over r^2}
{i\omega r\over v}
{\p\log c^2\over\p\log r}\delta S
,\label{pshigh}\\
{\Delta\over v^2}\left({\delta p_S\over p}\right)_{{\omega r\over v}\ll 1}&\sim&
{2\over r^2}{\p\log c^2\over\p\log r}
{\p\log\M\over\p\log r}\delta S\;\;\;{\rm if}\;l=0,\label{pslow0}\\
 &\sim&-{2\over r^2}
{\p\log\M\over\p\log r}{\p\log \over\p\log r}\left({v\over c^2r}\right)\delta S\;\;\;{\rm if}\;l\ge1.\label{pslow1}
\end{eqnarray}
In view of Eqs.~(\ref{pshigh}) to (\ref{pslow1}) and the advective-acoustic cycles described in F01, F02, two physical processes couple advected perturbations to the acoustic field:
\par (i) The gradient of temperature characteristic of the entropic-acoustic cycle is essential for spherical perturbations $l=0$ (Eq.~(\ref{pslow0}) and FT00) and high frequency perturbations (Eq.~(\ref{pshigh}) and F01).
\par (ii) Even in a isothermal flow (\ie $\p c^2/\p r=0$), non radial perturbations can excite acoustic waves in a vortical-acoustic cycle at low frequency (Eq.~(\ref{pslow1}) and F02). \\
The simple estimate in Eqs.~(\ref{pshigh}) to (\ref{pslow1}) shows that 
the strength of the excitator is comparable for radial and non radial perturbations, and that
its amplitude is highest near the lower boundary.\\
An efficient coupling between the excitator and the oscillator requires not only a strong amplitude of the excitator, but also a good matching of their spatial lengthscales. 
The wavelength of the excitator ($\sim 2\pi v/\omega$) is approximately a factor $\M$ smaller than the wavelength of the oscillator ($\sim 2\pi (c\pm v)/\omega$). This contrasts with the simpler case of black hole accretion, in which case the excitator and oscillator have comparable wavelengths near the sonic radius ($\M\to 1$).  The numerous oscillations of the excitator per acoustic wavelength should thus lead to a weak efficiency of the advective acoustic coupling in the inner regions where $\M\ll1$. The inner regions, however, are precisely the place where the amplitude of the excitator is highest, because the adiabatic gradients are strongest there. What is the net effect ? The choice of the lower boundary condition is crucial in answering this question.

\section{Boundary condition at $\rns$\label{sectbc}}

In order to separate the adiabatic effects from the non adiabatic ones, we choose to estimate the contribution of the adiabatic region by neglecting the acoustic feedback from the cooling layer as much as possible. The following assumptions are made at the lower boundary $\rns$: 
\par (i) acoustic perturbations propagating downward are perfectly reflected out 
($\omega\ll c_\stara /\rns$),
\par (ii) entropy and vorticity perturbations are freely advected below $\rns$ in the cooling region, where their coupling to acoustic waves is ignored. \\
Condition (ii) is equivalent to imposing that below $\rns$, entropy and vorticity perturbations cease to be source terms of the acoustic equation. The source term in Eq.~(\ref{diffpress}) is thus artificially damped by multiplying it by a smooth transition function $\Phi_\lambda$. The transition is assumed to take place over a length $\lambda$, comparable to the cooling length.  This damping of the source term can either be viewed as an ad-hoc damping of the entropy perturbation itself, independent of its frequency, or 
as a smoothing of the flow gradient responsible for the coupling.\\
The equations corresponding to these assumptions are obtained by matching the pressure perturbation
$\delta p$ for $r\ge\rns$ with the homogeneous solution $\delta p_0^0$ associated to Eq.~(\ref{diffpress}) for $r<\rns$ . 

\section{Eigenmodes of shocked accretion\label{secteigen}}

The Rankine-Hugoniot jump conditions are used to compute the 
perturbed quantities after the shock. These 
calculations are similar to those of Landau \& Lifshitz 
(1987, Chap.~90) extended to the case of a non uniform flow, or 
Nakayama (1992) extended to non radial perturbations.
For a perturbed shock velocity $\Delta v$ in the strong shock limit,
\begin{eqnarray}
{\delta v_r\over v_{\rm sh}}&=&{2\over\gamma+1}
\left({i\omega \rsh\over v_{\rm sh}}+{\gamma\over2}
{5-3\gamma\over\gamma-1}\right){\Delta v\over i\omega \rsh},\label{HCdv}\\
{\delta\rho\over\rho_{\rm sh}}&=&-{\gamma\over\gamma+1}{5-3\gamma\over\gamma-1}
{\Delta v\over i\omega \rsh},\\
{\delta v_{\Omega}\over v_{\rm sh}}&=&-{2\over\gamma-1}
{\partial\over\partial{\Omega}}{\Delta v\over i\omega \rsh}
,\\
{\delta p\over p_{\rm sh}}&=&-2{\gamma-1\over \gamma+1}
\left\lbrack {i\omega \rsh\over v_{\rm sh}}+
{5-3\gamma\over4}{\gamma^2+1\over (\gamma-1)^2}\right\rbrack
{\Delta v\over i\omega \rsh}.\label{HCdp}
\end{eqnarray}
The boundary value problem was solved numerically  for $l=0,1$. A broad range of unstable modes grow on a timescale comparable to a fraction of the  advection timescale. The radial mode is always the most unstable, closely followed by the non radial mode $l=1$.

\section{Evidence for the advective-acoustic cycle\label{sectwkb}}

Following the same method as F02, the discrete spectrum obtained in the boundary value problem is checked by computing, in two steps, the efficiency ${\cal Q}_{\rm adv}$ of sound production by the advection of an entropy/vorticity perturbation (without a shock), and the efficiency ${\cal Q}_{\rm sh}$ of entropy/vorticity production by an outgoing acoustic wave reaching a shock.
${\cal Q}_{\rm adv}$ is defined by:
\begin{eqnarray}
{\cal Q}_{\rm adv}&=&
\int_{\rns}^{\rsh} 
{1-\M^2\over2\gamma^2\M^2}
{\delta p_{0}^0\over p}\e^{i\omega\int_{{\rsh}}^r{1+\M^{2}\over 1-\M^{2}}{\d r\over v}}\nonumber\\
& &\times
\Phi_\lambda{\p\over\p r}
\left\lbrace
\left(1-{2\eta v\over i\omega r}\right){\omega^2c^2\over c_{\rm sh}^2\Delta}\right\rbrace
\d r.\label{qadvpdr}
\end{eqnarray}
The efficiency ${\cal Q}_{\rm sh}$ is obtained through a WKB approximation at high frequency:
\begin{equation}
|{\cal Q}_{\rm sh}|\equiv {1\over\M_{{\rm sh}}^{1\over2}}
{2\gamma\over 1+ {\gamma-1\over2\M_{{\rm sh}}}}.\label{defqshR}
\end{equation}
The global efficiency ${\cal Q}\equiv |{\cal Q}_{\rm adv}{\cal Q}_{\rm sh}|$ of the advective-acoustic cycle leads to a first estimate of the growth rate at high frequency: 
\begin{equation}
\omega_i={1\over\tau_{\rm tot}}\log {\cal Q},\label{estiwi}
\end{equation}
where $\tau_{\rm tot}$ is the total duration of the cycle (advection + acoustic). The exact resolution of the discrete eigenfrequencies was successfuly compared to the continuous WKB estimate (\ref{estiwi}). 

\section{Comparison with BMD03 \label{sectdiscus}}

Our stability analysis in an adiabatic flow should apply directly to the numerical simulations of BMD03. The authors recognized the existence of an advective-acoustic cycle similar to the one occuring in the vortical-acoustic instability of F02. Our linear study seems to agree qualitatively with their results concerning the mode $l=1$. Nevertheless, the mode $l=0$ is stable in their simulations, whereas it is the most unstable one according to our calculations, as well as in the work of HC92. This important difference may come from the "leaky" lower boundary condition of BMD03, which is different from ours even in the linear regime. 

\section{Conclusion}

The temperature and velocity gradients in the subsonic flow between the shock and the accretor is responsible for an efficient {\it linear} coupling between entropy/vorticity and acoustic perturbations. Most of the sound comes from the region close to the lower boundary of the adiabatic region, despite the fact that the wavelength of advected perturbations is much smaller than acoustic ones. The acoustic waves reaching the shock produce new entropy/vorticity perturbations, in an unstable cycle. The growth time is comparable to the advection time. The most unstable modes correspond to $l=0,1$ perturbations. The identification of the advective-acoustic cycle as the mechanism responsible for the instability was checked numerically through the calculation of ${\cal Q}_{\rm adv}$ and ${\cal Q}_{\rm sh}$.
Since the region contributing most efficiently to the instability is the vicinity of the lower boundary $\rns$, the role of cooling processes is at least crucial in determining the growth rate of the instability. The principle of a cycle between propagating and advected perturbations could be useful in interpreting the stability analysis of non adiabatic flows. Whether non adiabatic processes are partially stabilizing or even more destabilizing remains to be determined. The difficulties of 1-D numerical models in reaching an explosion (\eg Buras \etal 2003) suggests that cooling processes are indeed important enough to significantly stabilize the entropic-acoustic cycle at work for radial perturbations. \\
\\
{\it Acknowledgement.} The authors are grateful to T. Janka for his initial suggestion to study the effects of advective-acoustic cycles in the problem of core-collapse, and his permanent encouragements since then. Useful discussions with J. Blondin, A. Burrows and R. Chevalier are aknowledged.\\
 \\
{\bf References}\\
\\
Blondin, J., Mezzacappa, A. \& DeMarino, C. 2003. 
{\it Astrophys. J.}, {\bf  584}, 971--980 (BMD03)\\
Buras, R., Rampp, M., Janka \& H.-T., Kifonidis, K. 2003. 
astro-ph/0303171 \\
Foglizzo, T. 2002. 
{\it Astron. Astrophys.}, {\bf  392}, 353--368 (F02)\\
Foglizzo, T. 2001. 
{\it Astron. Astrophys.}, {\bf  368}, 311--324 (F01)\\
Foglizzo, T. \& Tagger, M. 2000. 
{\it Astron. Astrophys.}, {\bf  363}, 174--183 (FT00)\\
Houck, J.C. \& Chevalier, R.A. 1992. 
{\it Astrophys. J.}, {\bf  395}, 592--603 (HC92)\\
Landau, L. \& Lifshitz, E. 1987, Fluid Mechanics 6, Pergamon Press\\
Langer, S.H., Chanmugam, G. \& Shaviv, G., 1981. 
{\it Astrophys. J.}, {\bf 245}, L23--L26 (LCS81)\\
Nakayama, K. 1992. 
{\it Mon. Not. R. astr. Soc.}, {\bf  259}, 259--264\\
Saxton, C.J. \& Wu, K. 2001. 
{\it Mon. Not. R. astr. Soc.}, {\bf 324}, 659--684

\end{document}